# Magnetic, magnetocaloric and neutron diffraction studies on $TbNi_{5-x}M_x$ (M = Co and Fe) compounds

Arabinda Haldar[1], I. Dhiman[2], A. Das[2], K. G. Suresh[1,*] and A. K. Nigam[3]

[1]Department of Physics, Indian Institute of Technology Bombay, Mumbai- 400076, India

[2]Solid State Physics Division, Bhabha Atomic Research Centre, Mumbai- 400085, India

[3]Tata Institute of Fundamental Research, Homi Bhabha Road, Mumbai- 400005, India

*Abstract*

The effect of substitution of Co and Fe for Ni in $TbNi_5$ on the structural, magnetic and magneto-thermal properties has been investigated. Considerable enhancement of Curie temperature is observed with Fe substitution, whereas the increase is nominal in the case of Co. Neutron diffraction measurements reveal the redistribution of moments and site preference of substitutional ions in Ni $2c$ and $3g$ sites. In $TbNi_4Fe$, both Ni and Fe as well as Tb are found to carry moment while in the case of $TbNi_4Co$, mainly Tb carries the moment. Magnetocaloric behavior has been investigated from the magnetization and the heat capacity measurements. The magnetic and magnetocaloric properties are found to be strongly correlated in these compounds.

*Corresponding author (E-mail: suresh@phy.iitb.ac.in, Fax: +91-22-25723480).

## I. Introduction

$R$Ni$_5$ compounds (where $R$ is a rare earth ion) and their substitutional derivatives have been extensively studied because of their application in hydrogen storage technology and as adiabatic nuclear cooling agents. They are also attractive for fundamental studies by virtues of the distinct features of rare earth magnetism. Large magnetocrystalline anisotropy is present in these compounds, which originates from the crystal field interaction [1,2]. The fundamental interest stems from the fact that the magnetocrystalline anisotropy energy exceeds the exchange energy in these compounds and as a result crystal field plays an important role in determining the magnetic properties [2-6]. The fact that the indirect RKKY exchange is the predominant exchange interaction is another important feature of these compounds.

Compounds of this series with almost all rare earths have been subjected to various theoretical and experimental investigations [3,4,6]. Among these compounds, TbNi$_5$ has received the attention of many researchers. There are a few reports on the magnetic and neutron diffraction studies in TbNi$_5$ and its derivatives [7-12]. It is observed that in TbNi$_5$, the moment is carried by Tb mostly and that the Ni sublattice (almost completely 3$d$ shell) has negligible contribution to the moment. The ordering temperature ($T_C$) of TbNi$_5$ is 24 K. However, below this temperature, it is not a collinear ferromagnet, but is reported to possess a modulated ferromagnetic structure [7,8]. Several studies have also been made on this compound after replacing Ni with nonmagnetic elements like Al [9], Ga [10], Si [11], Cu [12] etc. With nonmagnetic substitution, the ordering temperature decreases. Decrease of Tb 5$d$ band polarization was ascribed to the diminution of magnetic contribution when replacing Ni by Al [9]. Anisotropy was found to play an important role in the magnetic properties in nonmagnetic substituted TbNi$_5$ compounds.

One of the important features of $R$Ni$_5$ series of compounds is the nonmagnetic nature of the transition metal sublattice, which leads to low $T_C$ values. In this respect, they are similar to the $R$Ni$_2$ compounds. The recent study of $R$(Ni,Fe)$_2$ series, in the context of magnetocaloric effect (MCE), shows various interesting results both from fundamental and application points of view [13,14]. With Fe substitution in place of Ni in HoNi$_2$ and TbNi$_2$, the ordering temperature was found to increase significantly. Though the MCE value decreases, the temperature variation of the isothermal magnetic entropy change shows double peak/broad peak behavior. Such an MCE behaviors is quite anomalous and is shown only by a very few materials. The reason for the anomalous MCE was attributed to the local anisotropy variations arising from the Fe substitution [13,14]. Since the crystal structure of $R$Ni$_5$ compounds is noncubic as compared to the $R$Ni$_2$ compounds, it is interesting to study the effect of possible local anisotropy variations resulting from partial Fe substitution in $R$Ni$_5$ compounds as well. It is also of interest to study the effect of partial Co substitution, since the magnetic moment of Co (in metallic state) is in between that of Ni and Fe. Therefore, we have selected the systems namely TbNi$_{5-x}$Co$_x$ and TbNi$_{5-x}$Fe$_x$ with $x$ = 0, 0.1 and 1. The structural, magnetic and magneto-thermal properties of these compounds have been studied in detail. We have also carried out neutron diffraction measurements at various temperatures in selected compounds of these two series.

## II. Experimental details

All the polycrystalline compounds, TbNi$_5$, TbNi$_{4.9}$Co$_{0.1}$, TbNi$_4$Co, TbNi$_{4.9}$Fe$_{0.1}$ and TbNi$_4$Fe, were prepared by arc melting of the stoichiometric proportion of the constituent elements of at least 99.9% purity, in a water cooled copper hearth in argon atmosphere. The resulting ingots were turned upside down and remelted several times to ensure the homogeneity. The weight loss was monitored at the end of the melting process and further characterization was performed only on samples with final weight loss less than 1%. The structural analysis of the samples was performed by collecting the room temperature x-ray diffraction patterns (XRD), obtained using Cu-K$_\alpha$ radiation. The refinement of the diffraction patterns was done by the Rietveld analysis using *Fullprof* suite program [15]. The lattice parameters were calculated from the refinement. The DC magnetization measurements, both under zero-field-cooled (ZFC) and field-cooled warming (FCW) conditions, in the temperature range of 1.8 - 300 K and fields up to 120 kOe were performed with the help of a vibrating sample magnetometer (Oxford). Some magnetization data and heat capacity data were collected in a Physical Property Measurement System (PPMS, Quantum Design Model 6500). The neutron diffraction patterns were recorded on the multi PSD based neutron powder diffractometer ($\lambda$ = 1.249Å) at Dhruva reactor in Bhabha Atomic Research Centre, Mumbai. The powdered samples were placed in a Vanadium can and attached to the end of a closed cycle helium refrigerator to vary the temperature between 20 and 300K. The refinement of the neutron diffraction patterns were carried out using *Fullprof* program.

## III. Results and discussion

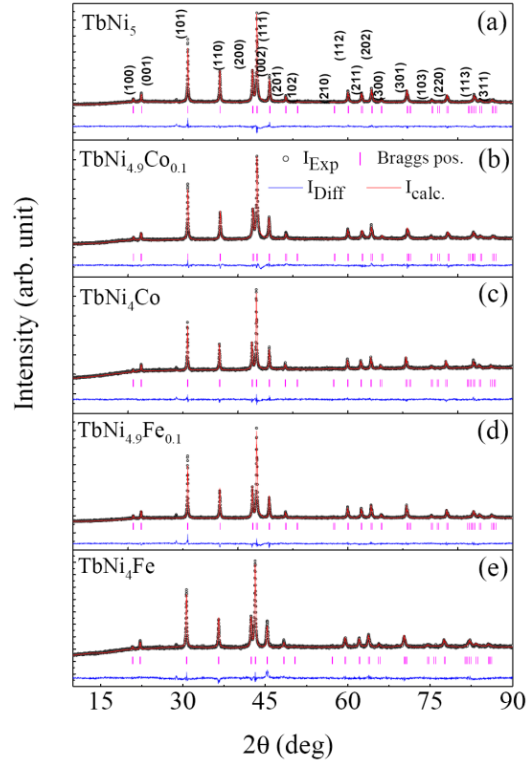

**Fig. 1.** Room temperature x-ray diffraction patterns, along with the Rietveld refined pattern of TbNi$_{5-x}$Co$_x$ and TbNi$_{5-x}$Fe$_x$ compounds. The plots below the Rietveld refined pattern show the difference between fitted and experimental data.

Fig. 1 shows the x-ray diffraction patterns of TbNi$_{5-x}$Co$_x$ and TbNi$_{5-x}$Fe$_x$ compounds. The Rietveld refinement of the XRD patterns confirms that all the compounds are single phase, crystallizing with MgCu$_5$ type hexagonal structure in the space group of *P6/mmm* (No. 191). The refined lattice parameters of all the studied compounds are given in Table I. Fe/Co substitution is found to increase both the cell parameters *a* and *c*. The *c/a* ratio too is found to increase with *x*.

**Table I.** Lattice parameters and the ordering temperatures in Tb(Ni,Fe/Co)$_5$ compounds.

| Compound | $a = b$ (Å) | $c$ (Å) | c/a | $T_C$ (K) |
|---|---|---|---|---|
| TbNi$_5$ | 4.8928 ± 0.0002 | 3.9631 ± 0.0002 | 0.8099 | 23 |
| TbNi$_{4.9}$Co$_{0.1}$ | 4.8949 ± 0.0002 | 3.9667 ± 0.0002 | 0.8104 | 29 |
| TbNi$_4$Co | 4.9026 ± 0.0002 | 3.9804 ± 0.0002 | 0.8119 | ~60 |
| TbNi$_{4.9}$Fe$_{0.1}$ | 4.8950 ± 0.0002 | 3.9684 ± 0.0002 | 0.8107 | 49 |
| TbNi$_4$Fe | 4.9219 ± 0.0003 | 4.0006 ± 0.0003 | 0.81281 | ~280 |

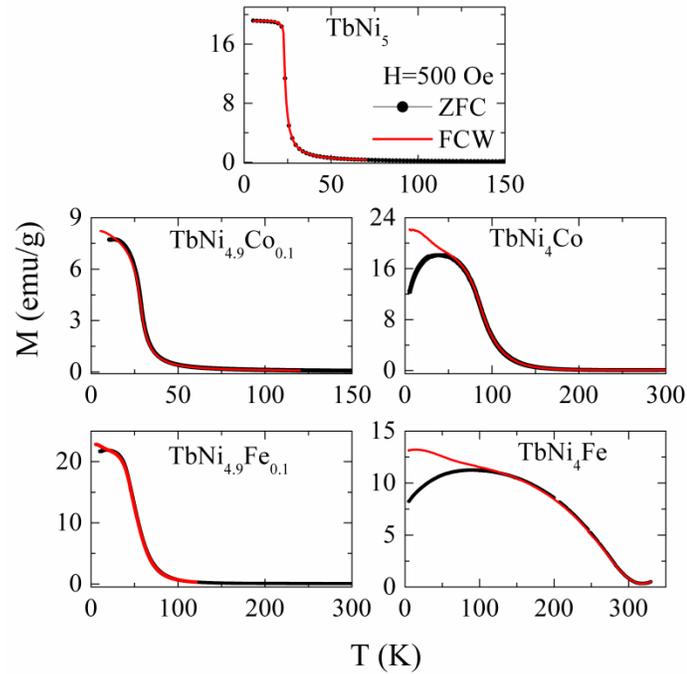

**Fig. 2.** Temperature dependence of the magnetization data (ZFC and FCW) of TbNi$_5$, TbNi$_{4.9}$Co$_{0.1}$, TbNi$_4$Co, TbNi$_{4.9}$Fe$_{0.1}$ and TbNi$_4$Fe in 500 Oe.

Figure 2 shows the temperature variation of magnetization ($M$) data of TbNi$_{5-x}$Co$_x$ and TbNi$_{5-x}$Fe$_x$ compounds under 'zero-field-cooled' (ZFC) and 'field-cooled' (FCW) conditions at $H$ = 500 Oe. The $T_C$ of all these compounds, determined from the first derivative of the ZFC plots, is listed in Table I. The magnetization exhibits a sharp transition at 23 K in the case of TbNi$_5$, which is in agreement with previous reports [16]. With substitution of Fe and Co, the transition becomes broader and the $T_C$ shifts to higher temperatures. The increase in $T_C$ is found to be significantly high in the case of Fe doping as compared to Co. The low ordering temperature of the parent compound is due to the absence of the direct exchange between the transition metal (*TM*) ions and the increase in the $T_C$ in the Fe/Co substituted compounds may be attributed to the increase in the *TM* (3*d*) -*TM* (3*d*) interaction. There may be an increase in the *R* (5*d*) –*TM* (3*d*) exchange interaction as well, which will also enhance the ordering temperature. Recent theoretical calculations on TbNi$_5$, TbNi$_4$Co and TbNi$_4$Fe confirm this proposition [17]. It can be seen from Fig. 2 that in the case of $x$ = 1, the transition is broad, as compared to that in $x$ = 0 or 1, for both Fe and Co. It can also be seen that the thermo-magnetic behavior of TbNi$_4$Co and TbNi$_4$Fe is different from others, at low temperatures. A downward turn of magnetization can be observed (with decrease in $T$), which indicates an increase in the coercivity or magnetic hardness at low temperatures. A similar behavior is also reported in LaNi$_4$Fe [18]. The ferrimagnetic coupling between Tb and *TM* moments may also contribute to this behavior.

From Fig. 2, it is also clear that the thermo-magnetic irreversibility between ZFC and FCW data is negligible in TbNi$_5$, TbNi$_{4.9}$Co$_{0.1}$ and TbNi$_{4.9}$Fe$_{0.1}$, whereas it is considerable in TbNi$_4$Fe and TbNi$_4$Co. This can be attributed to the additional magnetic hardness produced by random substitution of Fe/Co in an otherwise nearly nonmagnetic Ni subalttice [19]. For all the studied compounds, above $T_C$, the magnetic susceptibility ($\chi_{dc}$) could be fitted to the Curie-Weiss law $\chi = C/(T - \theta_P)$, where $C$ is the Curie constant and $\theta_P$ is the paramagnetic Curie temperature. The paramagnetic moment ($\mu_{eff}$) of TbNi$_5$ is found to be 11.5 $\mu_B$. It may be mentioned here that the theoretical value of paramagnetic moment for Tb$^{3+}$ is 9.7 $\mu_B$. Comparing the theoretical and experimental paramagnetic moments we can predict the presence of additional magnetic contribution to arise from the partial polarization of the transition metal sublattice. Based on the magnetic circular dichroism experiments in GdNi$_2$, Mizumaki *et al*. have suggested the presence of Ni moment of 0.2 $\mu_B$, which couples antiferromagnetically with the Gd moment [20]. To further understand this behavior, we have carried out neutron diffraction experiments, the results of which are presented later.

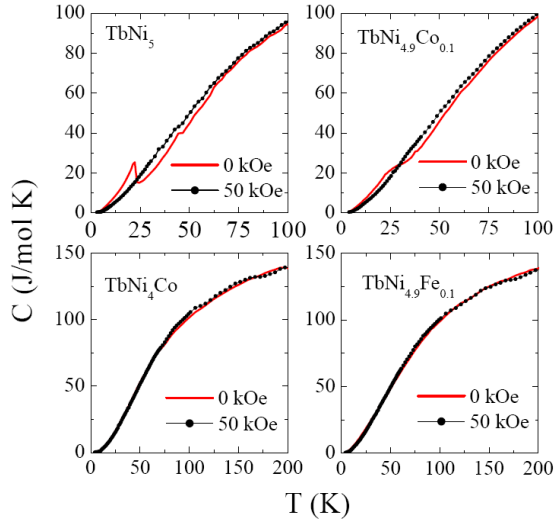

**Fig. 3.** Temperature dependence of heat capacity in TbNi$_5$, TbNi$_{4.9}$Co$_{0.1}$, TbNi$_4$Co and TbNi$_{4.9}$Fe$_{0.1}$ in zero and 50 kOe magnetic fields.

Fig. 3 shows the temperature variation of heat capacity (*C*) data for these compounds in zero and 50 kOe magnetic fields. In the parent compound, the *C vs. T* behavior shows a peak near its $T_C$ in zero field, but diminishes with the application of field. In the substituted compounds no such sharp peak is observed even in zero fled. This indicates that the sharpness of the transition decreases with substitutions/field and is in accord with the *M-T* data. Since the ordering temperature increases with Co and Fe substitutions, comparatively larger lattice and electronic part (at high temperatures) of the heat capacity would mask the weak peak due to magnetic contribution. A similar behavior has been observed in substituted *R*Ni$_2$ compounds also [13,14]. The nonmagnetic contribution arising from the lattice and electronic contributions was calculated from the zero field data. The magnetic contribution was calculated by subtracting the nonmagnetic contribution from the total heat capacity. The magnetic entropy ($S_M$) was then calculated from the magnetic heat capacity. In the case of TbNi$_5$, the temperature variation of the magnetic entropy shows saturation at high temperatures, with a saturated value of 26 J/mol K which is higher compared to the expected value of 21.3 J/mol K calculated from Rln(2*J*+1) [*J* = 6 for Tb$^{3+}$]. This may be attributed to the contribution arising from the partial polarization of Ni sublattice. On the other hand, all the substituted compounds show a non-saturating magnetic entropy behavior even at the highest temperature of 200 K. As the $T_C$ of all the compounds except TbNi$_4$Fe is much below this temperature, the non-saturation clearly suggests the presence of non-zero magnetic moment in the *TM* sublattice and the magnetic randomness associated it. The spin fluctuations of the *TM* moments must be the reason for this randomness. At 200 K, the value of $S_M$ for TbNi$_{4.9}$Co$_{0.1}$ is found to be about 24 J/ mol K, which again highlights the similarity of this composition with TbNi$_5$. The corresponding values in TbNi$_{4.9}$Fe$_{0.1}$ and TbNi$_4$Co are about 21 and 20 J/ mol K, respectively.

In order to further probe the magnetism in this series, neutron diffraction study has been carried out in TbNi$_4$Co and TbNi$_4$Fe. These two compositions were chosen since they have the maximum enhancement in the $T_C$. Neutron diffraction measurements have been carried out to investigate the distribution of the Ni and Co/Fe between the $2c$ and $3g$ sites and to find out the magnetic moment in each of the three sites. In contrast to TbNi$_5$ which exhibits an incommensurate antiferromagnetic structure [8], Fe and Co substitutions lead to a ferromagnetic behavior.

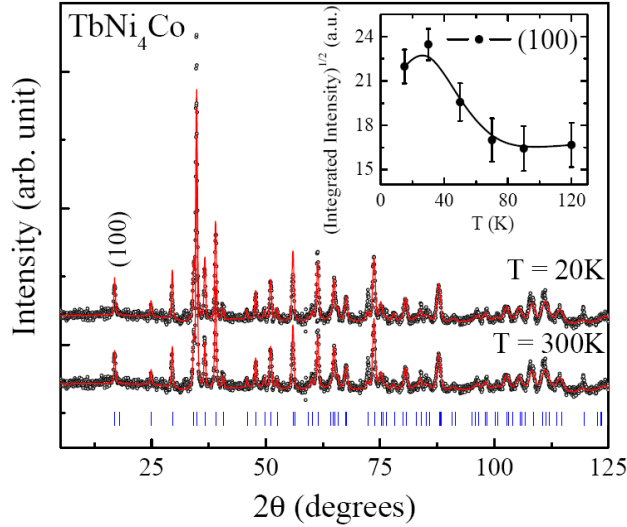

**Fig. 4.** Neutron diffraction patterns of TbNi$_4$Co at $T$ = 300 K and 20 K. Plots show experimental results with theoretically fitted line. The vertical tick marks indicate the position of the Bragg reflections. Inset shows the temperature variation of the square root of the integrated intensity of (100) reflection.

Figure 4 shows the neutron diffraction pattern obtained at 300 K and 20 K for $TbNi_4Co$. The Rietveld refinement of the pattern at 300 K shows that in this structure Tb occupies the 1$a$ (0 0 0) position, Ni occupies 3$g$ (1/2 0 1/2) and 2$c$ (1/3 2/3 0) sites. Co is distributed equally between the 2c and 3g site and is in agreement with the site occupancy for Co in $LaNi_4Co$ [21]. On lowering the temperature below 100 K, enhancement in the intensities of some of the low angle fundamental reflections is observed. This indicates the onset of long range ferromagnetic ordering in the system. No superlattice reflections were observed, which rules out any antiferromagnetic ordering, unlike in the case of $TbNi_5$ [8]. The ferromagnetic nature below 100 K is in conformity with the magnetization measurements, which shows an increase in the magnetization with lowering the temperature below 100 K. The best solution for the magnetic structure was arrived by placing the magnetic moment on Tb ion only. There may be a small transferred moment on the Ni and Co atoms, but this is below the level of detection in the present diffraction experiments. There exists a certain uncertainty in the results concerning the moment on Ni and Co. The structure factor of all the strongly magnetic reflections has contributions from all three, Tb, Ni, and Co ions. Therefore, from the present diffraction experiments it is difficult to conclude on the moment on Ni and Co. However, our choice of assigning moment only on Tb ion is guided by the absence of any magnetic ordering in $LaNi_4Co$ for $T > 5$ K. The magnitude of the moment on Tb is 6 $\mu_B$ at 17 K and is in agreement with the moment obtained from magnetization measurements. However, it is significantly lower than the expected moment of 9 $\mu_B$. In this regard, it is to be noted that the ab-initio calculations yields a similar Tb moment in TbNi$_4$Co. Moreover this calculation accounts for considerable moments on Co (~1 $\mu_B$) and Ni (0.3 $\mu_B$) [17]. The moment is found to be oriented along *c*-axis above 50 K. Below 50 K the moment is found to be tilted towards the *ab* plane, as indicated by the increase in intensity of (001) reflection below 50 K. Inset of Fig. 4 shows the temperature variation of the square root of integrated intensity of (100) peak, which is nearly identical to the *M-T* behavior shown in Fig. 2. This behavior was expected as the intensity is proportional to the square of the magnetic structure factor.

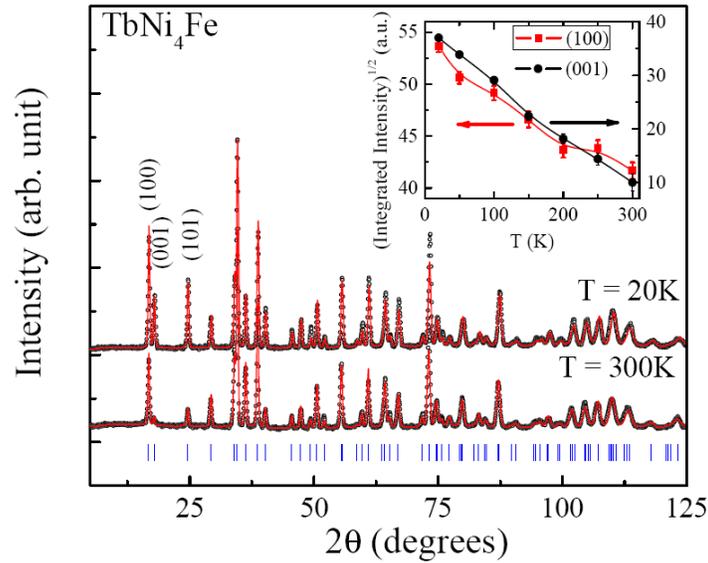

**Fig. 5.** Neutron diffraction patterns of TbNi$_4$Fe at $T$ = 300 K and 20 K. Plots show experimental results with theoretically fitted line. The tick marks indicate the position of the Bragg reflections. Inset shows the temperature variation of the square root of the integrated intensity of (100) and (001) reflections.

Neutron diffraction patterns have also been recorded in TbNi$_4$Fe compound at several temperatures between 300 K and 20 K. Fig. 5 shows the observed and the fitted data at 300 K and 20 K. At 300K, TbNi$_4$Fe exhibits ferromagnetic behavior, with $T_C$ just above 300 K as evident from the $M(T)$ plots. As against the Co case, in this case, Fe preferentially occupies the 3$g$ site. On lowering the temperature a significant enhancement in intensities of the low angle reflections (100), (001), (101), and (110) is observed (shown in the inset of Fig. 5). In the case of Co, such a large enhancement in (001) was not observed. This indicates a large tilt of the moments away from the $c$-axis. Moment is observed in Tb, Ni at 2$c$ and Ni at 3$g$ sites. The moment on Tb is 5.6 $\mu_B$, Ni at 2$c$ site has 0.1$\mu_B$ and Ni+Fe at 3$g$ site has 2.3 $\mu_B$. Similar moment values on the transition metal site have been reported from the theoretical studies carried out on LaNi$_4$Fe [22]. Ab-initio calculations on TbNi$_4$Fe have yielded almost similar values for Tb (6 $\mu_B$), Ni (0.4 $\mu_B$) and Fe (2.4 $\mu_B$) sites [17].

Comparing the neutron data on these two compounds, it can be seen that in TbNi$_4$Co, Co is distributed equally between the 2$c$ and 3$g$ sites and that the moment in the *TM* sublattice is very small. In contrast, in TbNi$_4$Fe, Fe prefers to occupy 3$g$ site and moment is distributed on both Ni and Fe ions. The *TM* moment is significantly larger in this case. Therefore, the neutron data clearly supports the larger increase of $T_C$ in TbNi$_4$Fe, as compared to that of TbNi$_4$Co.

The magnetocaloric properties of compounds under investigations have been studied in terms of isothermal magnetic entropy change ($-\Delta S_M$) using magnetization isotherms using the Maxwell's relation [23,24]

$$\Delta S_M(T, \Delta H) = \int_{H_1}^{H_2} \left( \frac{\delta M(T,H)}{\delta T} \right)_H dH \qquad (1)$$

For *M(H)* isotherms taken at different constant temperatures with discrete intervals, the above relation can be approximated to the following expression:

$$\Delta S_M \approx \frac{1}{\Delta T} \left[ \int_{H_1}^{H_2} M(T+\Delta T, H)dH - \int_{H_1}^{H_2} M(T,H)dH \right] \qquad (2)$$

Magnetocaloric behavior can be well parameterized from heat capacity measurement as a function of temperature in constant magnetic fields, $C(T)_H$. The entropy of a magnetic solid in zero field and in field can be expressed as [23],

$$S(T)_{H=0} = \int_0^T \frac{C(T)_0}{T} dT + S_0$$

and

$$S(T)_{H \neq 0} = \int_0^T \frac{C(T)_H}{T} dT + S_{0,H} \qquad (3)$$

where $S_0$ and $S_{0,H}$ are the zero temperature entropies. In a magnetic solid, these are the same (i.e. $S_0 = S_{0,H}$). Therefore, both the adiabatic temperature change [$\Delta T_{ad}(T)_{\Delta H}$] and $\Delta S_M(T)_{\Delta H}$ can be calculated as [24],

$$\Delta T_{ad}(T)_{\Delta H} \cong \left[ T(S)_{H \neq 0} - T(S)_{H=0} \right]_S \qquad (4)$$

$$\Delta S_M(T)_{\Delta H} \cong S(T)_{H \neq 0} - S(T)_{H=0} \qquad (5)$$

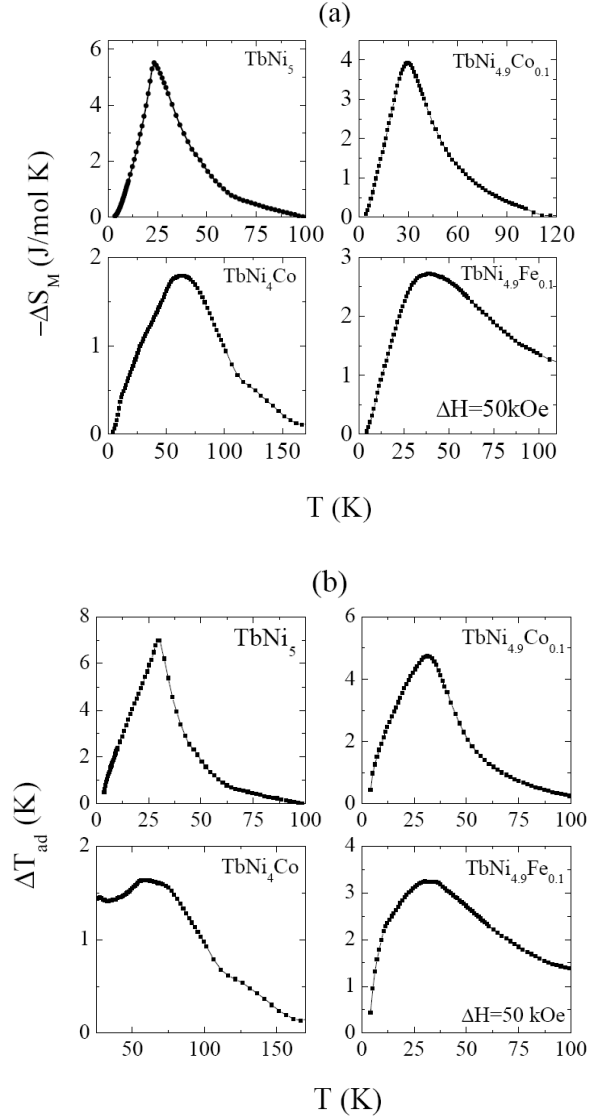

**Fig. 6.** (a) $(-\Delta S_M)$ vs. T plots for TbNi$_5$, TbNi$_{4.9}$Co$_{0.1}$, TbNi$_4$Co and TbNi$_{4.9}$Fe$_{0.1}$ compounds. (b) Adiabatic temperature change $(\Delta T_{ad})$ variation with temperature for TbNi$_5$, TbNi$_{4.9}$Co$_{0.1}$, TbNi$_4$Co and TbNi$_{4.9}$Fe$_{0.1}$ compounds. In both the cases, the field change is 50 kOe.

Figure 6(a) shows the $-\Delta S_M$ vs. *T* plots of the compounds for 50 kOe field change. It can be seen that $-\Delta S_M$ and $\Delta T_{ad}$ of TbNi$_5$ shows a maximum near $T_C$, as expected. For a field change of 50 kOe, the maximum values of $-\Delta S_M$ and $\Delta T_{ad}$ in TbNi$_5$ are ~ 5.6 J mol$^{-1}$ K$^{-1}$ and 7 K respectively, which are consistent with the theoretical prediction made earlier [6]. We observe that in Co-substituted compounds, $-\Delta S_M^{max}$ is smaller compared to that of the parent compound for the same field change. However, we get a broad peak at the highest concentration of Co. It can be seen that in TbNi$_{5-x}$Fe$_x$ compounds, the entropy change is even smaller, with a broad peak around $T_C$. Figure 6(b) shows the adiabatic temperature change in these compounds for various field changes. As in the case of entropy change, the adiabatic temperature change also decreases with Co/Fe substitution. It can be seen that the maximum in these plots also occur at temperatures close to the magnetic transition.

**Table II.** The maximum values of magnetic entropy change and adiabatic temperature change at $\Delta H = 50 kOe$ along with the ordering temperatures for the parent and the substituted TbNi$_5$ compounds.

| Compound | $T_C$ (K) | $-\Delta S_M^{max}$ (J/mol-K) $\Delta H = 50 kOe$ | $\Delta T_{ad}^{max}$ (K) $\Delta H = 50 kOe$ |
|---|---|---|---|
| TbNi$_5$ | 23 | 5.6 | 7.3 |
| TbNi$_{4.9}$Co$_{0.1}$ | 29 | 4.0 | 4.8 |
| TbNi$_4$Co | ~60 | 1.8 | 1.6 |
| TbNi$_{4.9}$Fe$_{0.1}$ | 49 | 2.7 | 3.3 |
| TbNi$_4$Fe | ~280 | 0.3* | --- |

*Calculated only from the *M-H-T* data

Table II gives the variation of maximum values of magnetic entropy change and adiabatic temperature change for $\Delta H = 50 kOe$ along with the ordering temperatures for the parent and the substituted compounds. It is noteworthy here that $\Delta S_M$ values calculated from the *M-H-T* data coincide with that calculated from the *C-H-T* data. The decrease in the magnitude of MCE with Fe or Co substitution is due to the reduction in the sharpness of the transition at the ordering temperature. We also find that the MCE behavior in these substituted compounds is more or less similar to that observed in Fe substituted $R$Ni$_2$ compounds [13,14]. As in $R$(Ni,Fe)$_2$ case, it is clear from Fig. 6 that the entropy change becomes considerable even at low temperatures ($T<T_C$). This implies that there is magnetic randomness in the zero field state even at such low temperatures. With the application of a field, the randomness decreases (magnetic entropy decreases), resulting in broad peaks in MCE, instead of a sharp peak at $T_C$. Since Fe/Co is substituted only partially, it is quite possible that there are local exchange and anisotropy variations. This gives rise to different local easy magnetization directions, thereby causing magnetic randomness, even below $T_C$. This also accounts for the broad magnetic transitions seen in the substituted compounds. Therefore, this study quite vividly reveals that the magnetic transition in the substituted compounds gets broadened due to the magnetic randomness and that the magnetic and magnetocaloric properties are strongly correlated. Since the magnetic moment of Co is not very much different from that of Ni, for low concentration of Co, we find that the magnetic and the magnetocaloric behavior are nearly similar to that of the parent compound.

## IV. Conclusions

Structural and magnetic properties have been studied in TbNi$_5$, TbNi$_{5-x}$Co$_x$ and TbNi$_{5-x}$Fe$_x$ compounds. The results show that the $T_C$ increases considerably with increase of Fe concentration, whereas the increase is nominal in the case of Co. The increase in the $T_C$ is attributed to the increase in *TM-TM* and *R-TM* exchange interactions which are quite negligible in the parent compound. In general, the neutron diffraction data corroborates the magnetization results. The analysis of the neutron diffraction data shows that both Ni and Co do not carry any significant moment in TbNi$_{5-x}$Co$_x$ compounds. On the contrary, the *TM* sublattice carries considerable moment in TbNi$_4$Fe. These conclusions are in reasonable agreement with the theoretical calculations. The magnetocaloric properties of compounds under investigation have been studied in terms of isothermal magnetic entropy change and adiabatic temperature change. MCE value is small in the substituted compounds compared to that of the parent compound, but we observe broad peaks in MCE in the former. The broad peak is attributed to the local variations in exchange and anisotropy, which causes magnetic randomness below $T_C$. In this respect, we find that there are similarities between Fe substituted $R$Ni$_2$ and the present compounds. The magnetic and magnetocaloric properties seem to be strongly correlated in these compounds.

## Acknowledgements

KGS and AKN acknowledge the financial support received from BRNS, DAE, Govt. of India for carrying out this work. The authors thank Dr. Pramod Kumar for his help in certain magnetization measurements.